\documentclass{article}

\usepackage{PRIMEarxiv}

\usepackage[utf8]{inputenc} 
\usepackage[T1]{fontenc}    
\usepackage{hyperref}       
\usepackage{url}            
\usepackage{booktabs}       
\usepackage{amsfonts}       
\usepackage{nicefrac}       
\usepackage{microtype}      
\usepackage{lipsum}
\usepackage{fancyhdr}       
\usepackage{graphicx}       
\graphicspath{{media/}}     
\usepackage{array}
\usepackage{multirow}
\usepackage{subcaption}
\usepackage{colortbl}
\usepackage{hyperref}
\usepackage{graphicx}
\usepackage{tabularx}
\usepackage{forest}
\usepackage{booktabs}
\usepackage{rotating}
\usepackage{pgfplots}
\usepackage{progressbar}
\usepackage{kvsetkeys}
\usepackage{adjustbox}
\usepackage{geometry}
\usepackage{pdflscape}
\usepackage[utf8]{inputenc}
\usepackage{tikz,pgfplotstable}
\usepgfplotslibrary{groupplots}
\usepackage{sparklines}
\usepackage{longtable}

\newcommand{\reza}[1]{#1}

\title{A Systematic Survey of Empirical User Studies of Unintentional Information Disclosure in Everyday Digital Interaction
}

\author{
  Reza Shahriari \\
  University of Florida \\
  Gainesville, FL \\
  \texttt{rshahriari@ufl.edu} \\
   \And
  Eric D. Ragan \\
  University of Florida \\
  Gainesville, FL \\
  \texttt{eragan@ufl.edu} \\
}

\begin{document}
\maketitle

\begin{abstract}

The exchange of personal information in digital environments poses significant risks, including identity theft, privacy breaches, and data misuse. 
Addressing these challenges requires a deep understanding of user behavior and mental models in diverse contexts.
This paper presents a systematic literature review of empirical user studies on unintentional information disclosure in usable security, covering 101 papers published across six leading conferences from 2018 to 2023.
The studies are categorized based on methodologies---quantitative and qualitative---and analyzed for their applications in various scenarios. Major subtopics, including data privacy, security in browsers, and privacy tools, are examined to highlight research trends and focal areas. 
This review provides details on topics and application areas that have received the most research attention.
Moreover, by comparing descriptive and experimental approaches, findings aim to guide researchers of strategies to mitigate risks associated with online everyday interaction.


\end{abstract}


\section{Introduction}
Online browsing and app usage are integral to both our personal and professional lives as people depend on various tools and applications to navigate the digital world.
Engaging in activities such as exchanging, submitting, and sharing information through emails \cite{wash2021knowledge,mayer2021now}, social media \cite{minaei2022empirical, amon2020influencing}, or other digital platforms \cite{emami2021understanding, yang2022you} is commonplace.
However, the online exchange of information comes with significant challenges and risks, particularly concerning privacy and security.
There is always a potential for unauthorized access and data breaches, which can lead to the release of personal and sensitive information, resulting in identity theft, financial loss, and other malicious activities.

To address online privacy and security issues, people adopt various protective behaviors \cite{youn2005teenagers, chen2017securing}.
These include downloading apps only from reliable and trusted sources and ensuring that the software they use meets high security and privacy standards. 
Additionally, they may seek to stay informed about potential threats by keeping up with security incidents affecting their families, friends, or the broader community, which helps them identify and avoid similar risks  \cite{mehrnezhad2022can,youn2005teenagers}.
Furthermore, they adjust their privacy settings to control who can access their information and posts, and they exercise caution before sharing personal details that malicious parties could misuse \cite{ebert2023creative, chen2017securing}.

Despite these efforts, only a limited number of users consistently familiarize themselves with or follow these risk prevention measures. 
Therefore, it is crucial to develop software and tools with well-established default security standards to minimize risks. 
Understanding how people engage with the digital world is essential for identifying specific challenges. 
The field of usable security and privacy (USP) helps identify these challenges at the intersection of human behavior and digital security \cite{sasse2005usable}.
USP helps security experts focus on critical risk areas, enabling more analysis of people's behavioral patterns and ultimately enhancing overall online safety.


While USP is a broad research area, this review specifically concentrates on the subtopic of online information disclosure in everyday digital interactions.
We use the term information disclosure to describe situations in which users, often unintentionally, reveal personal or sensitive data through routine online activities such as web browsing, email, or social media use. This focus reflects a consistent pattern in prior empirical studies, which examine how users make decisions about sharing, withholding, or managing data in digital environments. Studying this area is critical because users’ privacy and security are shaped not only by technical safeguards but also by individual choices about data sharing and software settings.

Additionally, people may struggle to grasp the nuances of potential information disclosure through different problems, vulnerabilities, and risks.
The dependencies on various applications and types of data further complicate this understanding.
Because of the need to understand user choice,
human-subjects research methodology is commonly employed to study user behaviors and thinking. 
Different choices of user studies, such as controlled experiments using hypothetical cases or surveys that reflect more realistic personal experiences, provide varying degrees of control and insight.
Experiments may offer more control but often rely on hypothetical scenarios, whereas surveys, though less controlled, may capture more authentic personal experiences.
Differences in user study types directly affect the knowledge gained, making it crucial to consider the overlap among methods applied, research subtopics focused on, and types of applications studied.

This review aims to highlight the value of a systematic literature review in organizing and assessing the research methods applied across different areas of information disclosure.
This will assist future researchers in selecting practical approaches for their studies and ultimately help lower users' online risks.
It provides an overview of areas that have received more attention and identifies key topics that could benefit from future work by providing evidence of what research methods have been most effected in different areas.
The review spans six leading research conferences in security and human-computer interaction in recent years (2018--2023), filtering from a total of 8185 papers to 101 papers given detailed analysis. 
We focus on empirical studies involving people, emphasizing the various methods used to conduct user studies.
Each method offers unique benefits and limitations.
Users have diverse experiences and perspectives that influence their disclosure decisions, which might vary significantly between laboratory settings and real-world contexts.
Personal past experiences differ from hypothetical scenarios, and there are distinct challenges in understanding risk when dealing with hypothetical cases versus real personal information.

\section{Background}
\subsection{Overview of Usable Security and Privacy}

USP aims to make security and privacy mechanisms both accessible and effective by balancing robust protection with user-friendly design. This necessitates a review of key areas where information is at higher risk of disclosure due to the nature of information exchanged or shared.
One common topic within USP is \textit{authentication}, which focuses on securing users' online accounts and data by verifying their identities \cite{ur2017design, bhana2020passphrase,liu2019reasoning}.
For instance, when a user logs into an email account, the authentication process ensures that the user is who they claim to be, typically through passwords or other verification methods.
In this context, a significant contribution to USP was the development and evaluation of a data-driven password meter, which measures the strength of passwords and provides users with actionable feedback to enhance password security \cite{ur2017design}.

Another crucial subarea in USP is the study of \textit{user behavior} for security and privacy.
This refers to the actions or decisions that individuals take that affect their privacy or data security.
For instance, Wu et al. \cite{wu2018your} explored how the explanations provided by browsers about private browsing modes shape users' misconceptions and beliefs.
Many users incorrectly assumed that private browsing would prevent geolocation tracking, protect against malware, eliminate advertisements, and stop tracking by websites and network providers.

Moreover, differences between expert and non-expert users of the Tor Anonymity Network emphasize the importance of understanding how Tor works to mitigate risks~\cite{gallagher2017new}. Experts have a technical understanding of Tor, while non-experts generally see it as a service without detailed knowledge of its operation.
This distinction is crucial for reducing the risks of deanonymization and ensuring user safety~\cite{gallagher2017new}.

Further investigation into why users follow or ignore computer security advice revealed key perception gaps drive these decisions.
For example, {Fagan et al.}~\cite{fagan2016they} found individuals who follow security advice tend to perceive greater benefits and risks in doing so, while those who ignore the advice see higher benefits in not following it.
This study highlights the crucial role of users' perceptions in their decision-making regarding computer security~\cite{fagan2016they}.

Research in USP also examines users' choices of \textit{security settings}, which are configurable options that help manage or enhance software features for the security and privacy of users and data. 
These settings include security warnings and indicators that should be easily accessible and understandable for users \cite{bravo2013your,felt2016rethinking,almuhimedi2014your}.
Bravo-Lillo et al. \cite{bravo2013your} explored the design of security-decision user interfaces, introducing attractors to draw attention to critical information in security dialogs, significantly reducing unsafe user actions.
On the other hand, Felt et al. \cite{felt2016rethinking} focused on browser security indicators, identifying deficiencies in existing designs through surveys of over 1,300 users. They proposed a set of new indicators: Secure for HTTPS, Not secure for HTTP, and Not secure for invalid HTTPS, based on extensive user testing and design considerations. 

Another subarea addresses the use of \textit{encryption}, the security method that involves converting information into a coded format to prevent unauthorized access. 
For instance, Abu-Salma et al. ~\cite{abu2017obstacles} conducted a study to explore obstacles to adopting secure communication tools. 
Their findings showed that fragmented user bases, incompatible tools, and users' lack of understanding of end-to-end encryption hinder its adoption. 
Additionally, Whitten and Tygar~\cite{whitten1999johnny} conducted a usability evaluation of PGP 5.0 \cite{garfinkel1995pgp}, a program designed for secure email communication through encryption and digital signatures. 
Their work~\cite{whitten1999johnny} highlighted the need for security-specific user interface design principles.
They also found that PGP 5.0 is not user-friendly for individuals unfamiliar with cryptography, leading to ineffective use of the security features.

Moreover, the topic of \textit{social engineering} focuses on manipulation techniques that trick individuals into revealing confidential information or providing system access to malicious actors.
As an example of work on this topic, Jagatic et al. \cite{jagatic2007social} demonstrated that attackers can easily and effectively exploit social network data to increase the success rate of phishing attacks. 
It was found that users are over four times more likely to fall, victim, if solicited by someone appearing to be a known acquaintance~\cite{jagatic2007social}.
Also, another study ~\cite{wash2018provides} explored the effectiveness of different phishing training approaches and found that the impact of the training significantly depends on who delivers it.
Facts-and-advice training was more effective when provided by security experts, while stories about phishing incidents were more impactful when shared by peers, showing the importance of the perceived origin of training materials~\cite{wash2018provides}.


Our review specifically concentrates on the USP subtopic of online information disclosure in digital interaction, which recognizes the need to protect most individuals who engage in online activities, as these users are often at risk of unintentionally disclosing information.
This focus encompasses a range of activities that threaten online users.
For instance, users often need to pay more attention to crucial security warnings, thereby missing out on vital information intended to protect them \cite{vance2019fog}.
Privacy concerns are also associated with using online proctoring software during virtual examinations, as these tools often require access to sensitive personal data \cite{terpstra2023online}. 
Furthermore, sharing personal information on social media platforms can expose individuals to various threats \cite{hasan2021your}, as can the misuse of private browsing features, with users frequently overestimating the level of privacy these functions afford \cite{habib2018away}. 
Additionally, using fitness devices that track and share users' health and location data can pose significant privacy risks if not properly managed and understood.
Each of these areas represents a critical aspect of online user behavior that necessitates careful consideration and study to enhance the safety and privacy of users in the digital world.

\subsection{Related Literature Reviews}

The literature survey we present in this paper adds to the body of knowledge on online information disclosure and usable security and privacy (USP) literature reviewed by others. Online information disclosure is a critical aspect of USP, as it encompasses the risks and behaviors associated with users' interaction and data-sharing practices. Our review aims to highlight the human factors influencing these interactions and how they align with the findings of previous reviews in the field.

Because our review focuses on human-subjects research, the review by Distler et al. \cite{distler2021systematic} is highly relevant.
Their review analyzed papers between 2014--2018 that involved human subject studies to study how researchers represent risk to participants, primarily through simulated or naturally occurring risk.
Their investigation also looked into the type of risk representation used across different research methods.
They found that papers with an experimental objective mainly employed simulated risk, while descriptive studies primarily used naturally occurring risk.
Various tools were used to represent risk to participants, including security/privacy-related tasks, prototypes, and scenarios.
Our review, however, differs by focusing on online information disclosure and covering a more recent period (2018--2023), allowing us to compare trends in USP methodology over time.

In addition, Iachello and Hong \cite{iachello2007end} highlight the critical role of HCI in privacy, emphasizing that many issues stem from user interactions with systems.
They call for standard privacy-enhancing techniques, effective analysis methods, and a comprehensive ``privacy toolbox''.
Similarly, Jacobs and McDaniel \cite{jacobs2022survey} find that non-expert misunderstandings of technology often lead to risky decisions and security breaches, stressing the need for user education.
Garfinkel and Lipford \cite{garfinkel2014usable} trace the evolution of usable security and privacy, emphasizing that poor usability often results in security failures, advocating for integrating usability and security in design.

\section{Methodology}




\subsection{Research questions}


Ensuring the security of individuals participating in online activities is crucial to prevent unintentional disclosure of personal information.
This includes addressing various online threats that can harm users.
In our review, we focus on online information disclosure in digital interactions, acknowledging the necessity to safeguard most individuals who engage in online activities, as they often risk unintentionally revealing information.
This area is chosen due to the significant influence of the user's decision-making process on data sharing and risk-taking.
Given the diversity in user experiences and perspectives, it is essential to employ human-subject research to capture this complexity and assess whether certain methods are more effective for specific subtopics.
This approach will help us understand if existing results indicate more appropriate methods for the research.

In this work, we sought to analyze research methodology used in USP addressing online information disclosure between 2018--2023. 
We conducted a systematic literature review of research papers published at top USP venues, filtering to a final set of 101 papers for full analysis.
Our analysis focuses on these research questions:
\begin{itemize}
\item \textbf{RQ1}: 
For research leveraging human-subjects methods, which topics and application areas have received the most research attention in the study of unintentional online information disclosure, and to what extent is this focus justified given their relevance to everyday digital interactions? Are there underexplored areas or overlaps among topics and applications that merit further investigation?

\item \textbf{RQ2}: 
What human-subjects research methods are commonly employed to study unintentional online information disclosure across different topics and application areas? What methodological details, such as participant pool, can guide future researchers in designing studies?

\end{itemize}

\subsection{Systematic Review}

This section outlines the methodology employed for conducting the literature review in this survey, providing an overview of its structure. 
Table~\ref{tab:filtering} provides an overview of the process.
Our systematic literature review approach involves four key phases: (1) publication venue selection, (2) search procedure, 
(3) filtering, and 
(4) detailed review.

\subsubsection{Publication Venue Selection}

The process began by identifying the most relevant peer-reviewed publication venues recognized for emphasizing USP papers, with a particular focus on those that highlight online information disclosure and the threats that users face in digital interaction with the online world.
Because USP papers are mainly presented in conference formats, we focused on top-tier venues that (i) are high-impact outlets in security, privacy, and HCI, (ii) consistently publish empirical, human–subjects USP work, and (iii) most directly intersect with explicit \emph{online} information disclosure. 

Balancing scope and feasibility, we selected three flagship conferences in the security and privacy domain: \textit{ACM Conference on Computer and Communications Security (ACM CCS)}, \textit{IEEE Symposium on Security and Privacy (IEEE S\&P)}, and \textit{USENIX Security Symposium (USENIX Security)}.
In addition, we included two privacy/USP–focused venues with high concentrations of human–subjects studies: the \textit{USENIX Symposium on Usable Privacy and Security (SOUPS)}, which is specifically tailored for USP papers and highly relevant to our study, and the \textit{Privacy Enhancing Technologies Symposium (PETS)}. 
Lastly, because usable security includes human-centered research, we also included the \textit{ACM Conference on Human Factors in Computing Systems (CHI)}, the premier HCI venue that routinely publishes empirical privacy and security work with human participants. 

We considered other strong venues (e.g., ESORICS, EuroS\&P, NDSS, CSCW), but excluded them after a scoping pass indicated substantially lower yield of human–subjects studies on online information disclosure relative to the screening effort.
We note this as a scope tradeoff and return to it in our limitations.
While these six venues are not the only outlets for high-quality USP research, they capture the highest–impact USP and HCI communities

\subsubsection{Initial Search Procedure}
The next stage of the review method applied term search and listing review to reduce the sample to the most relevant papers from the selected venues and time period.
In this stage, we employed two distinct approaches to account for differences in digital libraries and available search capability. 
In the first approach, we utilized the keyword search tools from the ACM Digital Library and the IEEE Xplore Digital Library to create our initial collection of possibly applicable USP papers for \textit{ACM CCS}, \textit{IEEE S\&P}, and \textit{ACM CHI}.

To select a set of papers covering user studies for online information disclosure, we first performed an initial filtering using keywords to cover a broad range of potentially relevant papers in usable security that we would later review for specific relevance to information disclosure. We used the keywords to select papers with at least one related term to usable security in addition to one related term to digital interaction in the title or abstract: (\textit{usable security} OR \textit{privacy} OR \textit{security}) AND (\textit{web} OR \textit{online} OR \textit{internet} OR \textit{email}).

We chose these terms because prior USP work shows most explicit online disclosure risks arise through web services, browsers, online communication, and email, not device-local or infrastructure-only contexts.
We intentionally did not include broad terms like \textit{IoT}, \textit{authentication}, or \textit{LLM} because they would expand beyond our scope and primarily return papers on mechanisms or platforms where disclosure is not the research focus.
In addition to searching through available online digital libraries, the search limitations of specific paper listings necessitated the second approach for some venues. 
Specifically, we manually reviewed each abstract for USENIX Security, SOUPS, and PETS, which lacked advanced search capabilities in the online digital libraries. 
Studies on social platforms were still captured because many abstracts/titles use \textit{online}/\textit{web}, and our manual venue passes (SOUPS/PETS/USENIX Security) recovered additional social-media–context papers that the keyword search may miss.
This allowed us to identify and include potential papers aligned with our research objectives.

\subsubsection{Initial Review and Filtering}

During our filtering process, we applied a multi-stage review of titles, abstracts, and full texts to eliminate papers that did not align with our research topic. This stepwise filtering is summarized in Table~\ref{tab:filtering}.

\paragraph{Stage 1: Title and keyword screen.}  
We first excluded papers whose titles made it clear they were unrelated to human-subjects research or online interactions. For example, papers describing purely technical mechanisms (e.g., ``Side-channel Attacks'') or system design papers without a user component were discarded. We also excluded non-archival items such as posters, extended abstracts.

\paragraph{Stage 2: Abstract review.}  
For papers that passed the first screen, we examined abstracts to ensure they (i) included empirical data from human participants and (ii) directly addressed disclosure-related behaviors, perceptions, or risks in online contexts. Examples of \emph{included} papers are those studying willingness to share data with mobile apps or expectations of privacy in video conferencing. Examples of \emph{excluded} papers are those focused solely on tool design without user evaluation, or those studying unrelated topics like cryptographic protocol design, authentication mechanisms, or IoT firmware security. 

\paragraph{Stage 3: Full-text eligibility.}  
In the final stage, we reviewed the complete texts to confirm that each paper (i) reported empirical human-subjects data (e.g., surveys, interviews, experiments, diary/log studies), (ii) investigated online information disclosure or user decision-making about sharing/withholding data, and (iii) provided sufficient methodological detail to extract study attributes. Papers failing these criteria, or tool-design papers without empirical user insights, were excluded at this step.

Our research focuses on studying the risks associated with digital interactions in online activities. By applying strict criteria, we ensure precision and relevance in our analysis.
This method allows us to comprehensively understand users' information leakage during their browsing or interaction experiences without their awareness.

\begin{table*}[ht]
\centering
\small
\begin{tabular}{|p{0.25\linewidth}|*{7}{>{\centering\arraybackslash}m{1.06cm}|}}
\toprule
\multirow{2}{*}{\centering Filtering Process} & \multicolumn{7}{c|}{Publication Venues} \\ \cline{2-8} 
 & ACM CCS & IEEE S\&P & USENIX Security & SOUPS & CHI & PETS & \textbf{Total} \\ \midrule
All published papers & 1053 & 760 & 1294 & 187 & 4381 & 510 & 8185 \\ \midrule
Potential papers after keyword searching and title review & 15 & 23 & 43 & 69 & 60 & 30 & 240 \\ \midrule
Potential papers after filtering with abstract review & 1 & 5 & 29 & 41 & 35 & 30 & 141 \\ \midrule
Included papers after final review & 0 & 3 & 20 & 26 & 25 & 27 & 101 \\ \bottomrule
\end{tabular}
\caption{Filtering selected papers in 6 publication venues between 2018-2023.}
\label{tab:filtering}
\end{table*}

\subsubsection{Detailed Review}
The paper selection and filtering process resulted in a final sample of 101 papers for detailed review.
To ensure a thorough analysis, the full text of all final papers was reviewed to gather information on the different research methods used, including surveys, experiments, interviews, and focus groups. 
Relevant notes, categorizations, and assigned attributes were documented in a spreadsheet to facilitate organization and analysis of this information.
Labels were created and updated iteratively over the course of the review.
If new labels emerged or new criteria were formed along the way, previously reviewed papers were re-reviewed to ensure consistency.
For each paper in the sample, we reviewed and recorded essential information such as the publication venue, topics, end-user context, application areas, research methods, related details, research questions, and research findings.
\reza{One researcher conducted descriptive coding of 101 included papers.
This was done to extract the general context of each paper. 
Then, two researchers iteratively reviewed the descriptive codes together and performed axial coding to select primary themes.
Next, both researchers then iteratively compared and discussed the resulting themes with taxonomies and related topics of other literature (with notable examples including {\cite{distler2021systematic}, \cite{jacobs2022survey}}).
Both researchers worked together to iterate on the themes and categories until reaching 
a consensus on the final set.}
Thus, the papers were distributed across different user experience and behavior topics, such as data privacy and information security, communication and privacy protection, browser-based security and privacy, privacy awareness, perceptions, and behaviors, privacy tools, online advertising, and tracking.

\section{Findings of Literature Survey}
\subsection{Research topic areas in online information disclosure}
To identify which topics have received more attention, understand which areas pose higher associated risks, and determine which areas require further research, we needed to develop a set of topic areas.
We based our choice of categories on a thorough analysis of the literature on online information disclosure and by coding the 101 papers in our sample.
Since no existing taxonomy exactly matched our criteria, we developed our own. This new taxonomy is grounded in existing literature, established taxonomies, and the specific intersections with our paper topic areas.

Each paper was categorized based on its contextual focus for the research into online security and information risk.
Following the descriptive coding of the papers, two researchers subsequently conducted a thematic review of the descriptive codes and reached a consensus on larger topic groups.
The resulting topics were aggregated into six predominant topic categories, as outlined in Table \ref{tab:topics}. 
\textit{Data privacy and information security} leads, capturing 30.7\% or 31 out of 101 papers of the research focus. 
Following this, \textit{communication and privacy protection} secures 21.7\% or 22 out of 101 papers, while \textit{browser-based security and privacy} represents 17.8\% or 18 out of 101 papers. 
Additionally, \textit{privacy awareness, perceptions, and behaviors} accounts for 12.9\% or 13 out of 101 papers, \textit{privacy tools} comprises 9.9\% or 10 out of 101 papers, and finally, \textit{online advertising and tracking} makes up 6.9\% or 7 out of 101 papers of the research distribution.


\subsubsection{Data privacy and information security}

The majority of the papers focus on users' mental models concerning the sharing of personal information with various entities. 
This topic investigates critical issues surrounding personal data, which is now commonly shared and stored online. 
It examines security risks linked to the posting of personal data on social media and poor data management practices, such as granting devices access to personal information like location data without understanding how it is managed.

For instance, a study found that many wearable activity tracker users significantly underestimate the number of third-party applications accessing their data, with most users having little understanding of these data-sharing processes \cite{zufferey2023revoked}.
In contrast, research on online proctoring by Terpstra et al. \cite{terpstra2023online} revealed that, under certain conditions, students found it acceptable to share data with their teachers, even when teachers were not directly involved in the proctoring process.

A critical question raised by this research is whether personal data that users have previously shared can ever be permanently deleted from the internet, particularly when shared with companies like Google and Facebook that hold vast amounts of user data. 
Investigations into users' understanding and expectations of online data deletion in social media and nonspecific contexts highlight users' desires for control over their data and the need for enhanced deletion mechanisms and preferences for data expiration \cite{minaei2022empirical,murillo2018if}.

Additionally, studies using responses to data transparency tools have explored users' concerns and perceptions about data collection by Google and Facebook.
These studies found that, while users appreciate the insight and control provided by these tools, they remain concerned about the amount of data being collected and shared \cite{farke2021privacy,{arias2022surprised}}.

\subsubsection{Communication and privacy protection}
This significant area focuses on protecting communications using various platforms, such as email, messaging, and social media.
It involves preventing potential risks that users may face, like phishing attempts through emails or the risks associated with using voice assistants or microphones in conferencing tools.

In email security, researchers have examined user decision-making and behaviors when confronted with phishing emails and malicious URLs.
For example, many studies concentrate on how different features can help users prevent these risks and understand their behaviors \cite{althobaiti2021don, liu2023understanding, zheng2022presenting}.
Interestingly, research found that untrained individuals often outperform phishing filters due to their familiarity and expectations regarding incoming emails \cite{wash2021knowledge}.
These studies highlight the importance of understanding user thought processes and actions.

The popularity of conferencing tools surged after the COVID-19 pandemic, leading to multiple studies on user behaviors and concerns when using these communication tools.
A common finding is that privacy is a top priority for users \cite{emami2021understanding}.
However, users often have limited control over their choice of conferencing tools, as these are frequently dictated by their company or colleagues.
Detailed investigations, such as one study on the usability issues of the mute button in video conferencing applications, found that while users perceive the mute button as a privacy control, various apps continue to access and even transmit background audio data \cite{yang2022you}.
This discrepancy between user expectations and actual app behavior underscores a significant usability problem that requires attention to improve user trust and privacy.

In the context of messaging, the primary concern for users is ensuring that their messages remain inaccessible to others.
This is achieved through end-to-end encryption, as highlighted in various research papers. These studies delve into users' mental models, behaviors, and misconceptions while using end-to-end encrypted platforms \cite{wang2023reporting}. 
Interestingly, it was found that visualizing encryption through icons and animations negatively impacted users' perception, while simple text was considered much more trustworthy \cite{stransky2021limited}.

\subsubsection{Browser-based security and privacy}
This topic analyzes the potential threats users might encounter while utilizing web browsers, addressing issues such as users neglecting warnings, installing ad blockers, and misconceptions about using private browsing tools like Tor.
The goal is to understand users' mental models and behaviors regarding web features and their privacy, aiming to design intuitive features or educate them on the risks associated with these actions to ensure their online safety and privacy.

Browser extensions are a popular feature among web users, and researchers have studied various aspects of them. For instance, one study examined users' general understanding and preferences regarding the data that extensions can access \cite{kariryaa2021understanding}.
Other studies have explored practical use cases, such as using browser extensions to prevent online tracking and addressing usability and breakage issues that users might encounter \cite{nisenoff2023defining, mathur2018characterizing}. 
These studies also explored whether users stopped using these extensions when faced with breakage.

More technically, the Domain Name System (DNS) is a fundamental web feature that allows users to type the name of a website instead of memorizing its numerical IP address. 
However, using an untrusted DNS can expose users to unauthorized data access or various types of attacks.
Several studies have focused on user awareness and concerns related to different encrypted DNS configurations and settings, as well as the use of an improved version called Protective Domain Name System (PDNS) and its adoption among users \cite{nisenoff2023user, rodriguez2023two}.
The research found that users mainly trust default configurations and do not customize their settings due to potential breakage. They also tend to adopt the settings used by their Internet Service Provider (ISP).

\subsubsection{Privacy awareness, perceptions, and behaviors}
Understanding users' behaviors, attitudes, and awareness levels regarding online privacy is crucial because it helps design more effective privacy measures and educational campaigns.
By comprehending how users perceive and react to privacy risks, developers and policymakers can create tools and guidelines that better align with users' needs and expectations.
This area aims to explore these aspects to comprehend users' expectations and perceptions of privacy. 
Although it shares similarities with other related topics, the primary focus is on the psychological aspect of users.

For instance, a study on the effect of social norms on privacy behaviors revealed that people tend to disclose sensitive data to avoid disagreeing with others \cite{rader2023data}. 
Individuals often imitate others' behaviors, especially when they perceive that others do not care about privacy.
Additionally, the literature highlights the importance of trust in privacy recommendations and the significant role of the source of these recommendations in building trust and encouraging the adoption of guidelines. 
People primarily educate those in their social circles, emphasizing the significance of trust in privacy-related matters \cite{gerber2022nerd}.

\subsubsection{Privacy tools}
Privacy tools encompass a range of features and tools designed to safeguard users' privacy online.
These tools offer users a secure and private connection to the internet, protecting them from online threats and prying eyes.
However, designing and developing these tools alone is not sufficient; it is equally important to understand users' mental models and scenarios when using these tools.
This understanding enables the creation of privacy tools that are more effective, user-friendly, and accessible.

One example of privacy tools is Personal Privacy Assistants (PPAs), which help users manage their privacy online. Research indicates that users prefer PPAs that can learn their preferences, offer high user involvement, allow vendor choice, and provide transparency around data disclosure \cite{stover2023investigating}. By incorporating these preferences, designers can develop more effective and user-friendly PPAs.

Additionally, nudging tools represent another type of privacy tool that can enhance users' privacy.
These tools use subtle prompts or reminders to encourage users to take actions that will improve their privacy and security \cite{frik2019promise, masaki2020exploring, story2020intent}.
For instance, a commitment nudge may remind users to install updates, back up their data, or enable two-factor authentication \cite{frik2019promise}.
Moreover, nudging tools have been used to influence the decision-making behaviors of adolescents using social network services (SNS) to avoid privacy and safety threats \cite{masaki2020exploring}.
Furthermore, nudging interventions have been utilized to promote the adoption of secure mobile payments \cite{story2020intent}.
By leveraging these tools, users can take proactive steps to protect their privacy and security online.

\subsubsection{Online advertising and tracking}

While relatively underrepresented in our sample due to the focus on user studies, online advertising significantly impacts privacy and data protection. 
Users' personal information and online behavior are analyzed to curate personalized ads, raising concerns about privacy and data protection.
Users often remain unaware of the extent to which their personal information is being tracked and used for advertising purposes. This indicates a need for further user studies to explore and address these concerns.
For instance, a study demonstrated the impact of hyper-personalized ads on users who had expressed negative emotions and were wary of privacy violations \cite{hanson2020taking}.
Another research highlighted the necessity for personalized and interactive ad-targeting explanations that address users' specific concerns \cite{lee2023and}.

Furthermore, the practice of personalized advertising can seriously affect privacy and security, as this information can be used to track and monitor users without their knowledge or consent.
Research has shown that users are not adequately protecting their data even when they feel confident about the type of data being collected \cite{gabriele2020understanding}. 
This uncertainty about data utilization can lead to significant privacy risks. 
For example, while "privacy zones" in fitness tracking apps aim to hide sensitive locations, they have proven to be ineffective \cite{mink2022users}.
This underscores the urgent need for better privacy protections and user education regarding the use and sharing of location data.

\begin{table}[h!]
\centering
\footnotesize 
\setlength{\tabcolsep}{3pt} 

{
    \renewcommand{\arraystretch}{1.5} 
    \begin{tabular}{>{\raggedright\arraybackslash}p{3.2cm}|r|l|r}
    \toprule
    \textbf{Topic} & \textbf{Number of Papers} & \multicolumn{2}{l}{\textbf{Percentage}} \\
    \midrule
    Data Privacy and Information Security & 31 & \color{black}\rule{1.8cm}{5pt} & 30.7\% \\
    
    Communication and Privacy Protection & 22 & \color{black}\rule{1.3cm}{5pt} & 21.7\% \\
    
    Browser-based Security and Privacy & 18 & \color{black}\rule{1.0cm}{5pt} & 17.8\% \\
    
    Privacy Awareness, Perceptions, and Behaviors & 13 & \color{black}\rule{0.9cm}{5pt} & 12.9\% \\
    
    Privacy Tools & 10 & \color{black}\rule{0.7cm}{5pt} & 9.9\% \\
    
    Online Advertising and Tracking & 7 & \color{black}\rule{0.5cm}{5pt} & 6.9\% \\
    \bottomrule
    \end{tabular}
}

\caption{Distribution of 101 papers across six topics related to privacy and security.}
\label{tab:topics}
\end{table}

\subsection{Application Areas}

Whereas the previously-discussed research topics addressed different subareas relevant to online information disclosure, our review also considered whether the studies addressed relevance for specific types of software applications.
While some studies addressed general techniques or concepts that were agnostic of specific applications, it is important to identify the different base application types that receive direct attention in empirical research.
Following our analysis, the resulting categories of \textit{application areas} were determined based on the specific platforms or categories of software where the research is applied.
Whereas most research subtopics are broader and not necessarily tied to specific use cases, the application areas are more concrete for types of software.
The purpose of identifying both topics and application areas was to understand the practical contexts in which researchers can gain more specific scenarios that demonstrate how research topics translate into practical applications.
They allow for cross-sectional analysis of how a topic is explored across different platforms.
For instance, if a leading topic is applied to only one application area, it may suggest a strong research community focus and a potential for saturation, or conversely, the need for more innovative solutions in this area.

Starting with the \textit{browser} application area that accounts for 21.7\% of papers and encompasses a wide range of activities, including browser extensions \cite{mathur2018characterizing, kariryaa2021understanding, frik2020impact}, cookie consent interfaces \cite{habib2022okay,bouma2023us}, user behavior, and concerns around web browsing privacy \cite{redmiles2020comprehensive, story2021awareness, smullen2021managing}, etc.
It is no surprise that every browser-based security and privacy topic paper explores \textit{browser} application areas.
However, only a quarter of papers in online advertising and tracking---the second most popular topic using this area---address this critical area.

\textit{Social media} applications were identified as another core application area with four topics identified in them, with communication and privacy protection being the most popular, accounting for 36\% of usage that investigated different types of privacy protection, such as sharing content \cite{amon2020influencing, hasan2021your}, account verification and trustworthiness \cite{xiao2023account, mink2022deepphish} to better to understand users' concerns and needs in the social media context.

Within the \textit{mobile} application domain, there are three topics that focused on these types of applications. An excellent illustration of this is using nudging techniques to encourage the adoption of secure mobile payments, as highlighted in \cite{story2020intent}.
As an example, \cite{wilkinson2020privacy} demonstrates the effectiveness of data exposure visualizations on mobile devices, further showing the use of this application's versatility across different topics.

\begin{figure}[h]
  \centering
  \includegraphics[width=1.04\linewidth]{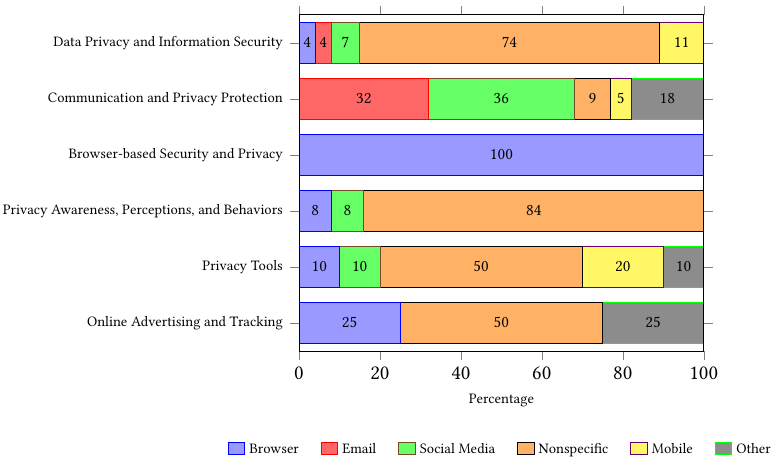}
\caption{Cross-tab of topics and application areas in percentage}
\label{tab:apptopics}
\end{figure}

Another identified category was \textit{email applications}, which were used mostly for studies of communication and privacy protection (see Figure \ref{tab:apptopics}). 
The review found 32\% of app areas focus on this topic and aim to provide a secure and safe space for users when communicating over email.
Under this area, some researchers have looked into adopting secure email (e.g.,~\cite{usman2023distrust}), while others have researched phishing email scenarios in this application area (e.g.,~\cite{althobaiti2021don, reinheimer2020investigation, wash2021knowledge, zheng2022presenting}).

In addition, many papers from the sample did not focus on a specific application area; rather, they explored broader topics that can be applied across various domains. 
We organized these papers under the \textit {nonspecific} application area because they are not specific to an application.
This category represents the most significant proportion of research papers from the sample, accounting for 41.5\%, as shown in Table \ref{tab:appareas}, demonstrating that a large amount of studies cover general concepts or techniques and often aim to be application agnostic. 
For instance, Sundar and Kim \cite{sundar2019machine} considered users' trust in humans vs machines, while Karunakaran et al. \cite{karunakaran2018data} studied users' understanding and opinions about data breaches.
Vance et al. \cite{vance2019fog} also researched the effect of security warnings that users tend to ignore, which can be applied in operating systems and browsers to show potentially malicious websites.

As seen in Figure \ref{tab:apptopics}, \textit{nonspecific} application area is substantial for four of the six topics, containing almost more than half of the papers.
This highlights the importance of exploring broader topics that can be applied across different domains and enables researchers to uncover fundamental concepts that can benefit multiple domains by studying \textit{nonspecific} application areas, leading to later in-depth analysis of needed areas.

During the application areas review process, we only received a few papers labeled as \textit{operating system} \cite{frik2019promise}, \textit{messaging} \cite{wang2023reporting, stransky2021limited}, and \textit{conferencing tools} \cite{emami2021understanding, yang2022you}. 
As a result, we decided to merge these and label them as \textit{other} for the application area.
As per our analysis, a significant portion (25\%) of the papers in this application area falls under online advertising and tracking. 
However, in three of the topics, there was no mention of this area.

\begin{table}[h!]
\centering
\small
\setlength{\tabcolsep}{2pt} 
\begin{tabular}{l|r|l|r}
\toprule
\textbf{Topic} & \textbf{Number of Papers} & \multicolumn{2}{l}{\textbf{Percentage}} \\
\midrule
Nonspecific & 42 & \color{black}\rule{2.62114cm}{5pt} & 41.5\% \\
Browser & 22 & \color{black}\rule{1.370672cm}{5pt} & 21.7\% \\
Social Media & 12 & \color{black}\rule{0.746088cm}{5pt} & 11.8\% \\
Mobile & 12 & \color{black}\rule{0.746088cm}{5pt} & 11.8\% \\
Email & 8 & \color{black}\rule{0.499064cm}{5pt} & 7.9\% \\
Other & 5 & \color{black}\rule{0.309684cm}{5pt} & 4.9\% \\
\bottomrule
\end{tabular}
\caption{Distribution of application areas in 101 papers}
\label{tab:appareas}
\end{table}

\subsection{Research Methods}

\subsubsection{Overview of Method Types}

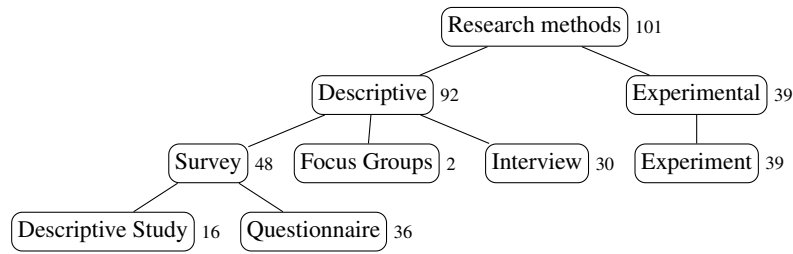
\begin{figure}
\centering
\scriptsize
\begin{forest}
  for tree={
    edge={-,>=latex}, 
    l sep+=0.1cm, 
    s sep+=0.44cm, 
    draw, 
    rounded corners, 
    align=center, 
    minimum height=0.1cm, 
    anchor=north, 
    fill=white, 
    font=\footnotesize 
  },
  [Research methods, label=right:101 
    [Descriptive, label=right:92 
      [Survey, label=right:48 
        [Descriptive Study, label=right:16] 
        [Questionnaire, label=right:36] 
      ]
      [Focus Groups, label=right:2] 
      [Interview, label=right:30] 
    ]
    [Experimental, label=right:39 
      [Experiment, label=right:39] 
    ]
  ]
\end{forest}
\caption{Research methods used in 101 papers}
\label{tab:tree}
\end{figure}

We categorized types of user research methods following criteria based on Lazar et al. \cite{lazar2017research}. 
Our categorization developed by applying the criteria to the findings in the paper sample to result in the following categories of study types: Descriptive and Experimental.
The approach for classifying method is similar to Distler et al \cite{distler2021systematic}.
However, we omitted the classification of relational studies due to the high dependence on the analysis method over the general method of study and the high frequency of overlap between relational research conducted following descriptive methods.
Our analysis found that a majority of papers (91.1\% or 92 of 101 papers) utilized \textit{descriptive research} methods, as demonstrated in Figure \ref{tab:tree}.
\reza{\textit{Descriptive research} aims to describe or identify behaviors or thoughts in a given situation \cite{lazar2017research}}. 
It often captures data from a natural setting and can serve as a foundation for further study. 
However, descriptive research only describes what is happening and does not provide insights into the causes behind observed relationships \cite{lazar2017research}.
Types of descriptive research methods found in our review included \textit{surveys}, \textit{focus groups}, and \textit{interviews}.
We discuss each of these methods in more detail in the next subsection below.

The review also found a large amount of experimental research, which focuses on comparison or testing hypotheses, often with the goal of testing causal relationships \cite{lazar2017research}. 
The review found 38.6\% or 39 out of 101 papers utilized experimental research methods. 
In order to establish cause-and-effect relationships, it is necessary to manipulate conditions and independent variables in a controlled environment and observe outcomes. 
However, it is important to note that causation and correlation are distinct concepts, and different conditions must be taken into account. 
Correlation refers to a statistical relationship between two variables, where changes in one variable are associated with changes in another.
However, correlation does not imply that one variable causes the change in the other. 
Causation, on the other hand, indicates that one event is the result of the occurrence of the other event; there is a cause-and-effect relationship.
Establishing causation requires more rigorous testing and evidence to show that changes in one variable directly result in changes in another, and that this relationship is not due to other confounding variables.
Therefore, simply conducting experiments in a controlled environment does not automatically establish causality.
We also discuss this in more detail in the following subsection below. 
By utilizing both descriptive and experimental research methods, researchers can produce more reliable and accurate results, which can benefit various fields of study.


\begin{table}[h!]
\centering
\small
\setlength{\tabcolsep}{3pt} 
\begin{tabular}{l|r|l|r}
\toprule
\textbf{Research Method} & \textbf{Number of Papers} & \multicolumn{2}{l}{\textbf{Percentage}} \\
\midrule
Survey & 48 & \color{black}\rule{3cm}{5pt} & 47.5\% \\
Experiment & 39 & \color{black}\rule{2.43cm}{5pt} & 38.6\% \\
Interview & 30 & \color{black}\rule{1.87cm}{5pt} & 29.7\% \\
Analyze Dataset & 12 & \color{black}\rule{0.74cm}{5pt} & 11.8\% \\
Focus Groups & 2 & \color{black}\rule{0.12cm}{5pt} & 1.9\% \\
\bottomrule
\end{tabular}
\caption{Research methodologies used in 101 papers }
\label{tab:methods}
\end{table}

\subsubsection{Survey}
Descriptive studies use surveys as the predominant method to collect information from individuals.
Surveys involve a set of structured questions that are designed to gather specific information from a target population. 
Surveys are often completed by individuals without the presence of a researcher, which can limit the depth and detail of the data collected. 
In very small cases, researchers may provide more in-depth insights or explanations and guidance to ensure that questions are properly understood.

The questions can be open-ended or closed-ended.
One of the strengths of surveys is their ability to gather a large number of responses from individuals quickly \cite{lazar2017research}.
This allows researchers to capture a wide range of perspectives and opinions on a particular topic.

The surveys were divided into two types: \textit{questionnaires} and \textit{descriptive study}.
In our sample, 75\% or 36 out of 48 papers that used the survey research method were structured \textit{questionnaires} to collect participant data.
These typically involve a set of predetermined questions that are administered to all participants, allowing for standardized data collection and analysis. 
For example, \cite{das2019typology} employed a questionnaire to gather data from 852 participants about the changes in their security and privacy (S\&P) behaviors, possible causes for these behaviors, and how they shared these behaviors with others.
The results revealed that social triggers, which involve interactions or observations of others, were the most common factor influencing S\&P behaviors, with a significant number of participants attributing changes in their behavior to interactions or advice from peers.

This example demonstrates how surveys can be used to collect data from a broad range of participants for age and gender diversity, which was crucial for this study. 
Surveys using questionnaires also allow data collection without the need for researchers to be present, which can help reduce unintentional researcher bias.

The remaining 25\% or 16 out of 48 surveys were classified as descriptive studies. 
\reza{Descriptive studies are a subcategory of survey studies and fall under the umbrella of descriptive research methods that are used to observe and record the characteristics of a phenomenon.
In these types of studies, participants engage in specific activities to observe and understand their behaviors and responses.
These studies are different than questionnaires that simply ask a set of questions to gather data.}
For instance, Zimmeck et al. \cite{zimmeck2023usability} conducted a study where participants were asked to go through a simulated browser setup process and make choices regarding various browser features, including the Global Privacy Control (GPC). 
The GPC has the potential to empower users to opt out of web tracking efficiently. 
The participants did not just answer questions--- they also actively engaged with a simulated interface to make setup choices that let researchers analyze usability.

\subsubsection{Experiment}
Experiments offer a controlled environment to explore cause-and-effect relationships, as explained earlier. The main objective of an experiment is to test hypotheses and determine which should be accepted or rejected based on statistical analysis \cite{lazar2017research}. For example, a study investigated the effectiveness of various measures in helping users identify phishing emails by involving 409 participants who were divided into two groups: one with a tutorial and one without \cite{reinheimer2020investigation}. The goal was to determine the effect of a tutorial on their ability to identify phishing emails. Researchers also exposed each group to different reminder measures, finding that measures based on videos and interactive examples performed best, with their effectiveness lasting for at least another six months.

By dividing the participants into different groups and exposing them to various measures, the researchers could identify which strategies were most effective. This controlled environment allowed for a thorough exploration of cause-and-effect relationships, and statistical analysis was employed to determine which hypotheses should be accepted or rejected.

\subsubsection{Interview}
Interviews provide valuable data that is difficult to obtain through surveys \cite{lazar2017research}. They allow researchers to explore thoroughly a problem and gather detailed responses through open-ended questions. Interviews encourage reflection and can reveal valuable insights that may not be captured in surveys. 

The use of interviews in research is invaluable for gaining insights into complex issues. For instance, a study interviewed 25 social media users to explore the relationship between the Fear of Missing Out (FoMO) and users' tendencies to compromise their privacy online \cite{take2022feels}. Through open-ended questions, the researchers gathered detailed responses about posting habits, joining and staying on platforms, leaving platforms, and perceptions of others' online habits and expectations. Another study used interviews to understand participants' experiences with shared accounts, interviewing 11 online and 14 in-person participants using a semi-structured approach that allowed them to express their thoughts freely \cite{obada2020burden}. Participants were presented with a categorized list of accounts and asked about their reasons for sharing, challenges when they stopped sharing, and their overall experience.
These studies demonstrate the power of interviews in gaining a deeper understanding of complex issues and uncovering insights that may not be captured through other means of data collection.

\subsubsection{Focus Groups}
Performing interviews is an effective means of data collection, but it can be time-intensive. 
This is because it necessitates individual meetings with each participant, which may extend to an hour or more per person. 
An alternative strategy is to utilize focus groups, where multiple participants can engage in collaborative discussions of their opinions, letting researchers gain a deeper understanding \cite{lazar2017research}.

As an example, Zhao et al. \cite{zhao2019make} conducted research on the online privacy awareness of children who are under 11 years old.
It aimed to investigate how well children can identify and deal with privacy risks that are related to their use of tablet computers. The research process involved 12 focus group sessions that consisted of 29 children.
The researchers used hypothetical scenarios featuring a cartoon character who experiences various online situations.
Subsequently, the children were asked to express their opinions on these scenarios, discussing what they and the character should do.

\subsubsection{Analyze Dataset}

Datasets were analyzed in around 11.9\% of the papers to identify patterns or trends and to gather more detailed information using other research methods. 
This step is often combined with other research methods, and it helps in finding themes to be able to use other research methods.
For example, Habib et al. \cite{habib2018away} conducted a study to explore the patterns and motivations behind users' engagement with private browsing modes in web browsers.
In the first step, they analyzed browsing data collected from more than 450 participants. These users had given their consent to have their daily computing behavior monitored through software.
This was followed by a survey to gain a better understanding of why people use private browsing for certain activities and whether they understand how it works.

\begin{figure}[h]
  \centering
  \includegraphics[width=1.05\linewidth]{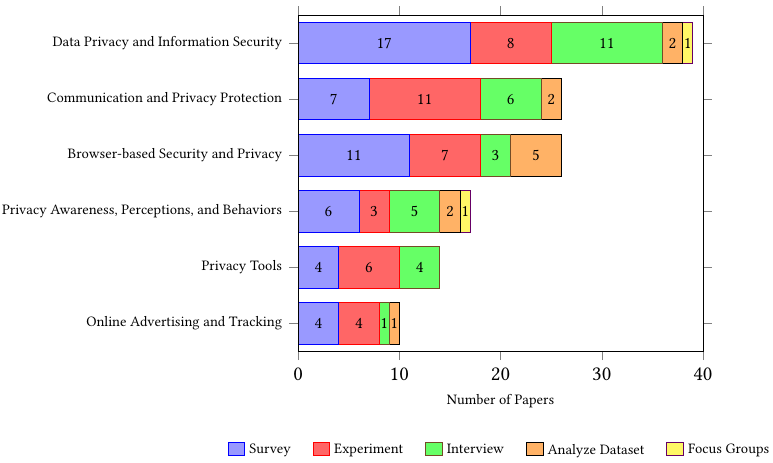}
\caption{\reza{Cross-tab of topics and research methods}}
\label{tab:studytopics}
\end{figure}

\subsection{Research Methods within Research Topics}

Our analysis also captured the intersection between type of research method and research topics.
The results (see Figure \ref{tab:studytopics}) show certain trends and preferences for approach within the different topics. 
Surveys are the most common method used, especially in data privacy and information security, where they are used around 44\% of the time. 
Experimental methods are popular in areas like privacy tools (43\%) and communication and privacy protection (42\%). 
They are important because they allow for the testing of tools and strategies in a controlled environment. 
Interviews are another frequently used method, particularly in data privacy and information security (28\%). 
They offer a more detailed and nuanced perspective, especially when individual experiences and perspectives are essential. 

Dataset analysis was used less often and usually only when relevant to the topic and appropriate data is available.
For example, it was more prominent in browser-based security and privacy (19\%). 
Focus groups are not commonly used in any of the topics, possibly because they do not provide the depth or specificity required.
Ultimately, researchers use specific methodologies tailored to the unique demands and specificities of each privacy-centric topic.

\subsection{Study Characteristics}

We also reviewed additional attributes of the different research methods found in the paper sample, as outlined in Table \ref{tab:methods}. 
This includes a detailed examination of critical factors such as whether the study was conducted in-person or online, synchronous or asynchronous (i.e., whether participants completed the study on in real-time with live communication with the researcher or if participants could independently complete the study without live researcher involvement), and whether it was performed in a group or individually.

\subsubsection{In-person vs. Online}
In terms of researcg methods, the Survey method was conducted entirely online, with all cases being carried out.
The Experimental method mostly relied on online methods, with 87.2\% of cases being conducted online and only 17.9\% being carried out in-person.
The Interview method showed a more balanced approach, with 66.7\% of cases being conducted online and 33.3\% in-person.
All the Focus Groups were conducted in-person.

\subsubsection{Synchronous vs. Asynchronous}
The survey method used in the study was conducted asynchronously, allowing all participants to engage at their convenience. On the other hand, the experiment method was mostly asynchronous, with 89.7\% of cases being asynchronous, whereas only 12.8\% were synchronous. The interview method was predominantly synchronous, with 93.3\% of cases being conducted in real-time, and only 6.7\% being asynchronous. Both of the focus group cases were conducted in real-time, making them 100\% synchronous.

\subsubsection{Group vs. Individual}

In the survey method, all studies were completed individually. Similarly, for the experiment method, the majority of cases (94.9\%) were individual-based, with only 7.7\% being conducted in a group setting. The interview method also showed a preference for individual sessions, with 90\% being conducted individually and only 13.3\% being group-based. As expected, focus groups were conducted in a group setting, representing 100\% of the 2 total cases.

\subsubsection{Number of Study Participants}

The data presents clear trends in the number of participants across different research methods, with quantitative methods showing the median participant numbers, as can be seen in Table \ref{tab:participants}.
Surveys, with a median of 393 participants, are particularly popular for gathering large-scale data. 
Surveys often aim to capture broad population-level insights, which necessitate higher participant counts to ensure statistical power and generalizability. 

Also, experiments follow closely with a median of 390 participants, reflecting their frequent use in controlled empirical investigations.
Experiments typically involve a controlled setup with a narrowly defined population to minimize variability and increase reliability. 
This focus often results in fewer extreme outliers compared to surveys, which might include very large participant counts (e.g., thousands in online surveys) for broad demographic insights. Consequently, while surveys can have a broader range, experiments maintain a more consistent participant count.

On the other hand, qualitative methods such as interviews and focus groups have considerably lower median participant numbers, 20 and 18, respectively.
This is primarily because these methods are inherently time-intensive and cannot typically be conducted asynchronously. 
Both interviews and focus groups involve direct, real-time interaction between researchers and participants, which limits the number of participants that can feasibly be accommodated within the constraints of time and resources.

Additionally, these methods often result in extensive, detailed qualitative data that require time-intensive analysis, such as thematic coding or transcription, further limiting their scalability. 
For focus groups, the need for coordination among multiple participants in a single session adds an additional logistical challenge, contributing to their smaller participant numbers.
This explains their less frequent application and lower participant counts, with focus groups being utilized in only two out of six topics.
In summary, the preference for quantitative methods like surveys and experiments is evident, as they dominate in participant numbers across most studies.

\newcommand{\scalemin}{0}
\newcommand{\scalemax}{10212}

\newcommand{\scalepoint}[1]{\numexpr((#1-\scalemin)*100/(\scalemax-\scalemin))\relax}

\newcommand{\minmaxbar}[6]{%
    \begin{picture}(100,10)%
    \put(\scalepoint{#1},10){\line(1,0){\numexpr\scalepoint{#2}-\scalepoint{#1}\relax}}%
    \put(\scalepoint{#1},8){\line(0,1){4}}%
    \put(\scalepoint{#2},8){\line(0,1){4}}%
    \put(\scalepoint{#3},8){\circle*{3}}%
    \put(\scalepoint{#1},4){\makebox(0,0)[b]{\tiny #4}}%
    \put(\scalepoint{#2},4){\makebox(0,0)[b]{\tiny #5}}%
    \put(\scalepoint{#3},12){\makebox(0,0)[b]{\scriptsize #6}}%
    \end{picture}%
}

\begin{table}[h!]
\centering
\footnotesize
\setlength{\tabcolsep}{3pt} 
\renewcommand{\arraystretch}{1.5} 
\begin{tabular}{l|c}
\toprule
\textbf{Research Method} & \textbf{Min-Max Bar} \\
\midrule
\textbf{Survey} & \raisebox{-6pt}{\minmaxbar{400}{15212}{2200}{29}{10212}{393}} \\
\textbf{Experiment} & \raisebox{-6pt}{\minmaxbar{9}{8994}{2200}{9}{4594}{390}} \\
\textbf{Interview} & \raisebox{-6pt}{\minmaxbar{9}{767}{380}{9}{67}{20}} \\
\textbf{Focus Groups} & \raisebox{-6pt}{\minmaxbar{7}{669}{400}{7}{29}{18}} \\
\bottomrule
\end{tabular}
\caption{Min-max bars representing the range of participant counts for each research method across all topics, with dots indicating the median values.}
\label{tab:participants}
\end{table}

\definecolor{lightgray}{gray}{0.85} 
\newcolumntype{L}{>{\raggedright\arraybackslash}X} 
\newcolumntype{C}{>{\centering\arraybackslash}X} 
\renewcommand{\arraystretch}{1.2} 
\begin{table*}[h!]
\centering
\tiny 
\begin{tabularx}{\textwidth}{L|C|C|C|C|C}
\toprule
 & \textbf{Survey} & \textbf{Experiment} & \textbf{Interview} & \textbf{Focus Groups} & \textbf{Analyze Dataset} \\
\midrule
\rowcolor{lightgray} 
\textbf{Data Privacy and Information Security} & 
\cite{arias2022surprised}, \cite{tolsdorf2022employees}, \cite{khan2021helping}, \cite{shen2021can}, \cite{van2020pets}, \cite{simoiu2019told}, \cite{latulipe2022unofficial}, \cite{ramokapane2021truth}, \cite{zhang2021did}, \cite{karunakaran2018data}, \cite{minaei2022empirical}, \cite{kroger2022personal}, \cite{khan2018forgotten}, \cite{zufferey2023revoked}, \cite{mayer2021now}, \cite{balash2022security}, \cite{stegman2022my} & 
\cite{hasan2018viewer}, \cite{sundar2019machine}, \cite{ul2023nuanced}, \cite{farke2021privacy}, \cite{terpstra2023online}, \cite{wermke2020cloudy}, \cite{sundar2020online}, \cite{liu2022your} & 
\cite{take2022feels}, \cite{murillo2018if}, \cite{mentis2019upside}, \cite{wilkinson2020privacy}, \cite{obada2020burden}, \cite{alqhatani2019there}, \cite{liu2022your}, \cite{khan2021helping}, \cite{shen2021can}, \cite{zhang2021did}, \cite{stegman2022my} & 
\cite{murillo2018if} & 
\cite{shen2021can}, \cite{van2020pets} \\
\textbf{Communication and Privacy Protection} & 
\cite{nthala2018informal}, \cite{emami2021understanding}, \cite{wash2021knowledge}, \cite{malkin2019privacy}, \cite{xiao2023account}, \cite{yang2022you} & 
\cite{liu2023understanding}, \cite{hasan2021your}, \cite{althobaiti2021don}, \cite{amon2020influencing}, \cite{mink2022deepphish}, \cite{reinheimer2020investigation}, \cite{zheng2022presenting}, \cite{stransky2021limited}, \cite{liu2022effects}, \cite{namara2021effectiveness}, \cite{bracamonte2022all} & 
\cite{reichel2020have}, \cite{wang2023reporting}, \cite{usman2023distrust}, \cite{ebert2023creative}, \cite{alqhatani2019there}, \cite{nthala2018informal} & 
 & 
\cite{xiao2023account}, \cite{yang2022you} \\
\rowcolor{lightgray} 
\textbf{Browser-based Security and Privacy} & 
\cite{maass2021effective}, \cite{nisenoff2023user}, \cite{kariryaa2021understanding}, \cite{mathur2018characterizing}, \cite{zimmeck2023usability}, \cite{smullen2021managing}, \cite{pugliese2020long}, \cite{rodriguez2023two}, \cite{utz2022privacy}, \cite{nisenoff2023defining}, \cite{habib2018away} & 
\cite{frik2020impact}, \cite{bouma2023us}, \cite{bernhard2019usability}, \cite{reeder2018experience}, \cite{maass2021effective}, \cite{thompson2019web}, \cite{habib2022okay} & 
\cite{huang2022users}, \cite{rodriguez2023two}, \cite{utz2022privacy} & 
 & 
\cite{huang2022users}, \cite{nisenoff2023defining}, \cite{thompson2019web}, \cite{habib2022okay} \\
\textbf{Privacy Awareness, Perceptions, and Behaviors} & 
\cite{hasan2023psychometric}, \cite{ray2020warn}, \cite{das2019typology}, \cite{riebe2023privacy}, \cite{redmiles2020comprehensive}, \cite{han2020price} & 
\cite{rader2023data}, \cite{langer2022open}, \cite{gerber2019investigating} & 
\cite{westin2021s}, \cite{redmiles2019should}, \cite{gerber2022nerd}, \cite{hasan2023psychometric}, \cite{ray2020warn} & 
\cite{zhao2019make} & 
\cite{redmiles2020comprehensive}, \cite{han2020price} \\
\rowcolor{lightgray} 
\textbf{Privacy Tools} & 
\cite{dutkowska2022and}, \cite{stover2023investigating}, \cite{story2021awareness}, \cite{namara2020emotional} & 
\cite{masaki2020exploring}, \cite{story2020intent}, \cite{kitkowska2020enhancing}, \cite{vance2019fog}, \cite{frik2019promise}, \cite{ebert2021bolder} & 
\cite{dutkowska2022and}, \cite{stover2023investigating}, \cite{masaki2020exploring}, \cite{story2020intent} & 
 & 
 \\
\textbf{Online Advertising and Tracking} & 
\cite{coopamootoo2022feel}, \cite{gabriele2020understanding}, \cite{hanson2020taking}, \cite{mehrnezhad2022can} & 
\cite{hanson2020taking}, \cite{tahaei2021deciding}, \cite{mink2022users}, \cite{lee2023and} & 
\cite{lee2023and} & 
 & 
\cite{mehrnezhad2022can} \\
\bottomrule
\end{tabularx}
\caption{Multidimensional summary of final 101 papers}
\label{tab:all}
\end{table*}

\section{Discussion}
\subsection{Analysis of Topics}


Understanding the distribution of research topics provides valuable insight into the focus areas and potential gaps in the field of usable security and privacy (USP).
This subsection examines the major themes of research and identifies trends, priorities, and areas needing further exploration.
As illustrated in Table \ref{tab:topics}, the distribution of papers across different topics shows that a large portion of papers (30.7\%) focuses on \textit{Data Privacy and Information Security}.
This likely reflects the critical importance of protecting personal data in today's world, where data breaches can have serious effects on both individuals and organizations.
The high number of studies highlights ongoing challenges like understanding how users handle passwords, share data, and spot phishing attempts.
It also indicates a need to develop user-friendly solutions that improve security without making things too complicated.

In contrast, topics like \textit{Privacy Tools} (9.9\%) and \textit{Online Advertising and Tracking} (6.9\%) have received less attention.
This difference raises questions about whether some important areas are being overlooked.
For example, \textit{Privacy Tools}—such as VPNs, encryption software, and privacy-focused browser add-ons—are key for protecting user data.
The lower research focus here might not mean these tools are less important; it could be that studying how usable they are and why people do or don't use them is challenging.
Many users find these tools complex or intrusive, which leads to low adoption rates.
This suggests we need more research to make these tools easier to use and to understand what prevents people from using them.

Similarly, the smaller number of studies on \textit{Online Advertising and Tracking} is notable given how common targeted advertising and tracking technologies are today.
These practices raise significant privacy concerns, like collecting personal data without consent and profiling users.
The lack of research might be because these tracking technologies are hard to study—they are always changing, and companies often hide them from users.
There's a pressing need to clarify how these technologies work and to assess their impact on user privacy.
This can help inform policies and lead to tools that help users manage their online footprints.


Figure \ref{tab:studytopics} shows trends in using research methods across different topics in privacy studies. 
For instance, \textit{surveys} are predominantly used in Data Privacy and Information Security, reflecting the method's strength in collecting large-scale quantitative data that can capture broad patterns and trends.
\textit{Surveys} allow researchers to gather data from a broad audience, making identifying general opinions and behaviors easier.
This approach is suitable for understanding the issues and concerns in \textit{data privacy and information security}.
Conversely, \textit{surveys} are among the least used methods in the study of \textit{privacy tools}. 
This indicates a reduced need for broad quantitative data, likely because the effectiveness and user interaction with \textit{privacy tools} may require more hands-on exploration, which \textit{surveys} are less equipped to handle.

On the other hand, \textit{experiments} are generally employed more than \textit{interviews} across various topics.
However, \textit{interviews} are utilized more extensively in the domains of \textit{data privacy and information security}, as well as \textit{privacy awareness, perceptions, and behaviors}.
Interestingly, these are also the only two topics where \textit{focus groups} where our review found examples of \textit{focus groups} employed, perhaps further highlighting the value of qualitative methods for understanding user thinking in these topic areas.
Using \textit{interviews} and \textit{focus groups} shows the need for in-depth, qualitative analysis to explore complex, subjective experiences and perceptions.
\textit{Focus groups}, in particular, bring the advantage of dynamic discussions, where participants can react to each other's perspectives, which is valuable for understanding deep or unspoken factors influencing privacy behaviors and perceptions.
However, these methods were used less frequently overall in the paper sample---likely because quality \textit{interviews} and \textit{focus groups} are time-consuming to organize and conduct, and it can be challenging to manage and analyze the often complex qualitative data that results from them. 

\subsection{Analysis of Study Methods}

Although \textit{surveys}, \textit{experiments}, and \textit{interviews} are popular study methods, \textit{focus groups} are less utilized with only 1.9\%, as shown in Table \ref{tab:methods}.
One reason for this could be the challenge of coordinating and managing communication among a diverse group of experts or individuals at a specific time.
However, \textit{focus groups} have limitations in the ability to assess usability because people may not know what they want or need \cite{krueger2014focus}. 
Additionally, online \textit{focus groups} are difficult to keep confidential and may not be representative of the average user, and are time-consuming and expensive.
However, it is essential to consider the benefits of \textit{focus groups} which can hgive researchers valuable information and a deeper understanding of users' mental models.
For example, \cite{murillo2018if} conducted \textit{focus groups} with seven data deletion experts that helped them to categorize different topics to be discussed with users and used their thoughts as a baseline to compare with \textit{interviews} that they did with users after that.
While \cite{zhao2019make} utilized \textit{focus groups} to understand children's mental models and privacy risks, which interactively gathered data that was not easily accessible through other research methods.

On the other hand, \textit{surveys} with a 47.5\% usage rate are commonly used in various research studies due to their ease of conduct and the reduced need for collaboration and control. 
\textit{Surveys} can save time and allow for the collection of a vast amount of information from a broad range of people.
In survey research, most of the studies conducted tend to involve questionnaires rather than descriptive studies, according to Figure \ref{tab:tree}.
While questionnaires are more accessible to create and administer, it might be a good idea to consider having users perform a task or a hypothetical scenario instead and then gather information from them such as in \cite{ramokapane2021truth} which aimed to understand how users provide false information and tell privacy lies online by asking participants to imagine buying a movie ticket as part of a hypothetical task.

Descriptive studies could result in better insights since questionnaires often rely on self-reported data, which can be flawed and inaccurate, but by observing users' behaviors, researchers can get an understanding of how users interact and how their behaviors may change over time, leading to better research outcomes.

Although \textit{surveys} and \textit{experiments} are commonly used methods in many fields, analyzing datasets in conjunction with user studies is not widely utilized across all topics covered in this review. 
Given the potential benefits of \textit{analyzing dataset} in understanding the problem and context before conducting a user study, this can have a significant impact.
For instance, it is interesting to note that \textit{analyzing dataset} is predominantly used in \textit{browser-based security and privacy}, with 19\% of researchers using it to qualitatively analyze users' online comments about Chrome's notification \cite{huang2022users}.
This data was then used to conduct \textit{interviews} later.
However, it is worth noting that none of the researchers utilized this method in \textit{privacy tools}.
This is a missed opportunity, as different types of online data from \textit{privacy tools} like password managers, VPNs, or privacy nudges may be available for analysis, which could yield valuable insights.

Building on these observations about study methods and their applications, it is useful to compare our findings with prior reviews, such as Distler et al.'s research, to contextualize trends and differences over time. 
It is important to note that our review primarily focuses on topics related to online information disclosure, while Distler et al. \cite{distler2021systematic} took a broader approach to papers in the field of USP employing human-subjects studies during the period preceding our own review.
Although both reviews analyzed different study methods and topics, their objectives were different. 
Their analysis concentrates on the strategic choices made by researchers in representing risk in their studies, which can impact the study's design and yields, particularly in fields where understanding risk perception and behavior is essential.
In contrast, our review aimed to categorize and compare specific research topics and methods, conducting cross-comparisons across different application areas and directly examining varied case studies to understand methodological patterns in the context of online privacy and security behaviors.

According to Distler et al.'s research, which reviewed papers from 2014--2018, \textit{experiments} were the most commonly used method at 35\%, followed by \textit{interviews} and \textit{surveys} at 13\% and 12\%, respectively. 
However, in our paper, from 2018 to 2023, \textit{surveys} were found to be the most commonly used method at 47.5\%, followed by \textit{experiments} and \textit{interviews} at 38.6\% and 29.7\%, respectively.
Both \textit{surveys} also noted that analyzing datasets and conducting \textit{focus groups} were less commonly used methods.
Focused groups were less utilized because they are less effective for creating realistic or controlled representations of security and privacy risks.
Unlike experiments, which allow for precise manipulation of risk scenarios, or surveys, which can capture naturally occurring behaviors, focus groups often rely on shared discussions that may not align with the study’s need for individual or measurable responses to risk.


Distler et al. \cite{distler2021systematic} found that studies with descriptive methods often involved naturally occurring or mentioned risks, highlighting how risks were studied in prior research.
This context helps frame the methodology and focus of current findings, particularly in understanding the relevance of risk representation to study objectives.
This is because descriptive methods usually provide less opportunity for risk simulation and are better suited for evaluating real-life risks or mentioned risks using \textit{interviews} or \textit{surveys}.
Our analysis found that \textit{data privacy and information security} emerged as the leading topic with the most papers, which often focuses on descriptive methods due to its emphasis on understanding users' behavior when sharing personal data, where naturally occurring risks are more relevant.
In contrast, experimental methods are more prevalent in areas such as \textit{communication and privacy protection} and \textit{privacy tools}, as these topics benefit from controlled environments to simulate risks and test user responses to tools and messaging platforms.
This aligns with Distler et al., who highlighted the importance of experimental setups for studying risk simulation effectively in these contexts.


\subsection{Application Areas within Different Topics and Methods}

Understanding the application areas of research topics and methods bridges theoretical insights with practical implementations.
This section examines how various empirical studies on online information disclosure are applied across specific software platforms and contexts.
While 41.5\% of papers from the reviewed sample focused on nonspecific application areas, as shown in Table \ref{tab:appareas}, diving into specific sections that provide more details for each context might also be essential.
\reza{Differences in types of applications and technologies will logically translate to different forms of privacy risks, as demonstrated in Figure \ref{tab:apptopics}.
While studies with more generalized concepts are undoubtedly useful in covering widely applicable insights that can apply to many application contexts, the general approach also risks missing out on key insights into the unique risks posed by more specific applications.} 

It is therefore critical to also conduct targeted, application-specific research, and the results of our review suggest more attention to key areas may be warranted.
For example, despite the increasing use of mobile apps, common adoption of smartphones, and expanding use of mobile applications for services, Figure \ref{tab:apptopics} shows research of mobile apps received less attention during the sampled period. 
Mobile apps tend to collect and use a significant amount of personal data, but they do so in different ways compared to non-mobile apps.

Similarly, social media is a significant part of modern life for many people, that received relatively small amount of focus overall---and the analysis did not yield any direct examples of intersection between specific \textit{social media} applications and the topic of \textit{online advertising and tracking}.
Understanding the specific risks within these contexts is essential if we want to develop targeted privacy solutions.

\reza{
These example demonstrate the value of considering the focus of different application areas, as a broad coverage with both specific and nonspecific applications is important for developing more nuanced privacy protections.
Within application categories, the focus on browser applications was high (21.7\% or 22 papers) compared to other specific areas such as social media and mobile.
As almost every person uses these applications, it is crucial to pay more attention to them as much as other areas and mitigate the risks associated with their usage.
It is worth noting that email is another application area that had relatively lower overall focus in sampled user studies.
This is possibly due to the longer history of email usage in daily life, and research advancements in phishing algorithms can also filter out many untrustworthy emails by taking advantage of non-expert users \cite{wash2021knowledge}.
However, attention to this area is still necessary as email is an essential communication tool for most people.
}

\section{Limitations}
\reza{The analysis presented here is based on only a limited number of papers in USP field involving human-subject studies.
At USP, many areas use human subjects, which can provide a large sample size, as demonstrated in studies such as Distler et al. \cite{distler2021systematic}, to evaluate the impact of different empirical methods.
However, this paper narrows the focus to online information disclosure, primarily concerning scenarios where human decision-making and understanding can affect privacy.
As we are confident that this topic is important to evaluate, the existing literature and taxonomies lacks a categorization of topics in this area due to its narrow focus. Therefore, our sub-topic extraction might not align with past literature but can be used as a starting point in this area.
As USP papers published in journals were limited, we did not consider them a major publication venue and were missing from our sample.}

\reza{As described previously, our review concentrates on the methods used to study online information disclosure.
This includes cases where users may not be fully aware of the risks, which often leads to unintentional information disclosure. As a result, we have excluded topics such as encryption, authentication, and IoT/mobile security, which directly address these types of risks.
Nonetheless, we acknowledge that some empirical research in these areas may have been inadvertently omitted from our review.}

\section{Conclusion}
We systematically reviewed 101 papers in USP-related research dealing with online information disclosure as seen in Table \ref{tab:all}, which allowed us to categorize each paper based on research methods, topics, and application areas.
We looked into the intersection of different categorizations, such as using study methods in topics and application areas across topics, to reveal how researchers utilize various methods to gain insight across other areas.
The review explored various methods used by researchers to gain insight across different areas and concluded that descriptive research methods were preferred over experimental methods. 
Surveys in questionnaire structure were particularly favored over descriptive formats that required users to perform a task.
Surveys in questionnaire structure were particularly favored over descriptive formats that required users to perform a task.
Additionally, privacy tools and email in application areas seem to receive less attention, indicating that there is potential for improvement in the development of tools that can enhance privacy protection, particularly in the context of phishing scenarios.
The researchers should pay attention to the sensitive topic of online advertising and tracking much more.
This issue has the potential to put many aspects of privacy and security in danger. It can allow attackers to infer sensitive information and even execute location-related attacks.

\bibliographystyle{unsrt}  
\bibliography{references}

\end{document}